\documentclass{aastex}

\begin{document}

\title{FOURIER ANALYSIS OF GAMMA-RAY BURST LIGHT CURVES: 
       SEARCHING FOR DIRECT SIGNATURE OF COSMOLOGICAL TIME DILATION}

\author{Heon-Young Chang}

\affil{Korea Institute For Advanced Study \\
207-43 Cheongryangri-dong, Dongdaemun-gu, Seoul, 130-012, Korea}
\email{hyc@kias.re.kr}

\begin{abstract}

We study the power density spectrum (PDS) 
of light curves of the observed 
gamma-ray bursts (GRBs) to  search for 
a direct signature for  cosmological
time dilation in the PDS statistics with 
the GRBs whose redshifts $z$'s are known.
The anticorrelation of 
a timescale measure and a brightness 
measure is indirect evidence of its effect.
On the other hand, 
we directly demonstrate that a time dilation 
effect can be seen in GRB light curves. 
We find that timescales tend to be shorter 
in bursts with small redshift, 
as expected from cosmological time-dilation 
effects, and we also find that there may be 
non-cosmological effects constituting to this correlation. 
We  discuss its implication on interpretations of 
the PDS analysis results.
We put forward another caution to
this kind of analysis when we statistically 
exercise with GRBs whose $z$ is 
unknown.

\end{abstract}


\section{INTRODUCTION}

Cosmological objects should not only be 
redshifted in energy but also extended
in time because of the expansion of the universe. 
Time dilation is a fundamental property of 
an expanding universe. 
There have been interests in cosmological 
time dilation as an observational experiment 
where its effect is large, such as high-redshift 
quasar or supernova observations.
In order to measure time dilation in light 
curves of the cosmological objects, 
it is necessary to find a way of
defining the timescale and of characterizing 
the timescale of variation.
The autocorrelation function has also been 
widely used for this purpose.
A number of groups have looked for time 
dilation in quasar light curves and reported 
 their successes (e.g. Hook et al. 1994),  
but there seem other opinions 
on the detection (e.g. Hawkins 2001).
A more direct observation of time dilation 
has come from the measurement 
of the decay time of distant supernova light 
curves and spectra \citep{l96,r97}. 
The results so far published are very convincing 
and strongly imply that time dilation has 
been observed. Another cosmological object where 
one would expect to see a time dilation effect is 
the observed gamma-ray bursts \citep{pac92,pi92}.

Observations taken by the BATSE instrument aboard 
the {\it Compton Gamma Ray  Observatory} 
have identified more than a few thousand 
gamma-ray bursts (GRBs) and 
shown that their angular distribution 
is highly isotropic implying that GRBs are 
at a cosmological distance \citep{p99}. 
Observations of the afterglow of GRBs
enable us to establish the fact that GRBs 
are indeed cosmological \citep{mao92,meegan92,pi92,met97}. 
If GRBs are at cosmological 
distances then the burst profiles 
should be stretched in time due to 
cosmological time dilation 
by an amount proportional to the redshift, $1+z$. 

Without knowing their redshifts, different groups 
\citep{n94,m96,che97a,che97b,lp97,ds98,lbp00} 
have investigated the correlation of the duration 
of bursts and the burst brightness
in order to look for a signature of time dilation. 
There have been a number of claims by groups 
working on GRBs that time 
dilation is seen in the stretching of peak-to-peak 
timescales \citep{n94,n95,lp97,ds98,lbp00}.
The expected redshift range of order unity would result in 
a time-dilation factor of a few while the burst 
durations cover a large 
dynamic range from tens of milliseconds to 
thousands of seconds \citep{fish95}. 
Therefore, a time-dilation effect can only 
be detected statistically.
One of the most serious limits of previous 
works is that 
inferences are all {\it indirect} and possibly misleading 
since the redshifts of most GRBs are unknown. 
\citet{n94,n95} searched for time-dilation 
effects by dividing the bursts into groups 
based on their peak count rate and comparing 
some measure of burst duration with peak count 
rate. They have claimed that brighter 
bursts had shorter durations than dimmer ones 
and that the difference between the 
average durations of bright and dim bursts was 
consistent with a time-dilation factor 
of about 2. 
If bursts were standard candles, dimmer bursts 
would be time-dilated more than brighter 
bursts, by a dilation factor 
$(1+z_{\rm dim})/(1+z_{\rm bright})$, where  $z_{\rm dim}$ 
and  $z_{\rm bright}$ are the redshifts 
of dim and bright bursts. 
However, finding cosmological time dilation signature in
light curves of GRBs is disputed.  For instance,
\citet{m96} finds no time dilation in BATSE 
using the aligned peak test, 
and \citet{b94} has warned that an intrinsic 
burst luminosity 
function could easily produce similar effects. 
Even if there is a correlation between the 
duration measure and the brightness measure
of the bursts, it is not clear that the 
argument can be inverted to provide convincing 
evidence for the existence of time dilation. 
Questions have been raised as to whether or 
not the time stretching 
that is found is due to the intrinsic correlation 
between pulse width and burst 
brightness for bursts drawn from a volume-limited 
sample \citep{bra94,bra97}.
\citet{ym94} also noted that relativistic 
beaming in either Galactic halo or cosmological 
models can produce flux-duration 
relationships that might be consistent with the 
reported effects.
\citet{wp94} suggested 
a way to distinguish between anticorrelations 
between flux and duration produced by 
cosmological time dilation and those produced 
by a decrease in burst density with 
distance, which is needed in a local extended 
halo model if  the luminosity 
function is independent of distance.

It is clear that despite the numerous works 
published on the subject, time dilation of GRBs 
remains controversial. Here we present
direct results on this topic which differ from 
those of previous works 
in two important ways.
Firstly, we analyze the Fourier power spectra 
of a sample of 
GRB light curves to look for such an effect. 
It provides a significant advantage over 
other methods, which
is relatively easy to interpret.
All the timescales of GRB variability are expected 
to show the effect of time
dilation. We do not require to isolate one 
particular timescale to fit, which may cause
artificial results. Secondly,
we use light curves of the GRBs whose redshift $z$ 
is known so that we are 
able to infer the time dilation effect directly. 
Statistical significance is reduced 
because of a small size of GRB data sets. 
Nonetheless we have a direct measure of time 
dilation, if it were, since we
use the GRB light curves for $z$-known samples. 
The number of the GRBs whose $z$ is measured is
increasing steadily, and it is worth while to 
attempt directly confirming time dilation effects with
the GRBs with redshift information.

\section{PDS OF GRB LIGHT CURVES}

We have used light curves of GRBs from the 
updated BATSE 64 ms ASCII 
database\footnote{\rm $ftp://cossc.gsfc.nasa.gov/pub/data/batse/$}. From this
archive we select the light curves of the GRBs 
whose redshifts are available.
We list up the GRBs used in our analysis with 
BATSE trigger numbers and 
the reported redshifts in Table 1. We divide 
our sample into two subgroups so
that we separate near and far GRBs.
We calculate 
the Fourier transform of each light curve of GRBs
and the corresponding power density spectrum (PDS), 
which is defined by the square of
the Fourier transform of the light curve.  
Before averaging the calculated PDSs in each subgroup, 
we normalize GRB light curves by setting their peak 
fluxes to unity. We compare the slopes obtained by 
the linear fits as it is without a time dilation correction
with those after rescaling the individual GRB light curve
to factor out $1+z$.
We have repeated 
the same process for the light curves of four 
different energy bands. 

In Figure 1, we show the averaged PDSs for the two 
subgroups of the GRBs
divided by the redshifts. 
Open triangles 
and squares represent the far
GRBs ($z \ga 1.5$) and the near GRBs ($ z \la 1.5$), 
respectively. 
For the far GRB subgroup power in lower frequencies 
is high, and for the near GRB subgroup power is 
concentrated in high frequencies. This is exactly what one 
may expect if light curves of
GRBs are lengthened due to cosmological 
time dilation. 
Instead removing the individual Poisson noise of a burst
from the individual PDS at high frequencies before averaging,
we attempt power-law fits in the limited range, i.e., $-1.6 < \log f < 0$.
The lower bound is determined in  that the deviation from the 
power law begins due to the finite length of bursts.
The upper bound is given such that the Poisson noise 
becomes dominant. In fact, this is the range where the Poisson
noise can be negligible and consequently 
the subtraction of the noise can be ignored, 
as seen in Figure 2 of \citet{bel98}. 
Poisson noise of the 
time bin becomes important only
at high frequencies, $f \ga 1$ Hz. Besides, it is the 
range where the simple power-law can be applied \citep{bel98}.
Dashed lines and
solid lines are the best fits obtained by least 
square fits for the far
GRBs  and the near GRBs, 
respectively. 
Before the fitting the averaged PDS is smoothed
on a scale of $\Delta \log f =0.2$. 
The slopes and standard deviations
obtained by the linear fit is summarized in Table 2.
Exact values of the obtained slopes are subject to the range
used in the fitting process. However, the trend is
hardly affected, that is, the subgroup of far GRBs results in
the steeper slopes than the one of near GRBs.
For all channels, the subgroup of the GRBs 
with higher redshifts result 
in exclusively steeper slopes compared with that with lower redshifts.
The slopes of the channel 4 show that 
peaks in higher energy bands are narrow in general.

To see the effects of time dilation, we rescale 
the time interval  of individual 
GRB light curve by a factor of $(1+z)^{-1}$, 
where $z$ is the redshift of the individual GRB. 
This should remove the effect of time dilation, 
that is, the difference of slopes in the two subgroups 
resulting from cosmological time dilation. 
This manipulation has the effect of shifting 
the contributions of all GRBs to the range
 of higher frequencies.  Resulting slopes of 
the fits are shown in Figure 2 and summarized in Table 2. We 
note that the removal of the ($1+z$) 
factor makes discrepancies of slopes in two 
subsamples reduced indeed, but differences
still marginally remain. 

\section{DISCUSSION}

Claiming time dilation in light curves of GRBs 
with the anticorrelation 
of a timescale measure and a
 brightness measure has several difficulties.
One difficulty is that this effect is correct only 
for standard candle sources with 
a standard duration, which we have evidence that 
it is not necessarily true 
\citep{kcy01,cy01}.
A broad luminosity function and/or an intrinsic 
spread in 
the durations could smear out the signature. 
Another possible difficulty with 
this anticorrelation is that it could be 
mimicked by intrinsic properties 
of the sources \citep{bra94,bra97,ym94,wp94}. 
An additional complication is that an intrinsic 
redshift 
of the time profiles from higher energy bands to 
lower energy bands 
may be present \citep{fb95}, which would 
bleach the cosmological signature.

We investigate the correlation between redshifts 
and timescale measures using available GRB data
with known $z$. Unlike past indirect searches for 
cosmological time dilation, we use the GRBs
whose $z$ is known at the expense of statistical 
significance.  
Diverse time scales shown in GRB light curves may 
result from 
cosmological time dilation of bursts, or from 
intrinsic properties of burst sources. 
The correlations among pulses within individual 
bursts give a measure of the intrinsic effects, 
while the correlations among bursts could result 
from both intrinsic and cosmological 
effects. We find that timescales tend to be shorter 
in bursts with small redshift, 
as expected from cosmological time-dilation effects, 
but we also find that there may be
non-cosmological effects constituting to this  
correlation. 
The implication of our analysis is that light 
curves of the observed GRBs show both 
intrinsic and cosmological effects. It is shown 
from Figures 1 and 2 that removing time 
dilation effect indeed reduces discrepancies in 
trend of time scale in the two subgroups
divided by the redshifts. However, it is not clear 
that difference remained after taking into account 
a dilation effect is due to other effects pointed 
out previously (e.g., Brainerd 1994, 1997;
Yi \& Mao 1994). 
Because of  the small number of data, 
it is inconclusive that these imperfect
corrections require other explanations other than 
cosmological time dilation. The amount of 
observed stretching {\it may not} be
the value expected from cosmological time dilation 
alone \citep{hmk96,mm96}. Challenging
questions then are whether one may extract information on
intrinsic properties of individual GRBs or whether one 
may distinguish a cosmological model by an analysis of
the slope of the observed PDSs of GRBs.

Another important implication of our study  should be 
pointed out. 
\citet{bel98} applied the Fourier transform technique 
to the analysis of light curves of long GRBs. 
They claimed that, even though individual PDSs 
were very diverse the 
averaged PDS was in accord with a power law of 
index $-5/3$ over 2 orders of 
magnitude of a frequency range, and that 
fluctuations in the power were 
distributed according to the exponential 
distribution.
With due care, the analysis of such kind 
may yield valuable information of 
the central engine of GRBs \citep{pan99,cy00}. 
However, the averaged power law index and the 
distribution of 
individual power should be corrected first
in terms of  a time 
dilation effect before making any
physical points out of the results of the PDS analysis. 
We have followed similar procedures for the total sample 
as \citet{bel98} did and 
obtained the slopes 
$-1.6074\pm0.105, -1.6423\pm0.099, -1.6876\pm0.094,~{\rm and}~-1.2190\pm0.086 $,
from the channels 1 to 4, respectively, which are close 
to the reported value $-5/3$ indeed. 
However, these slopes become flatter
when time dilation correction is made before the analysis, that is, 
$-1.5253\pm0.112, -1.5184\pm0.114, -1.506\pm0.122,~{\rm and}~-1.222\pm0.087 $,
from the channels 1 to 4, respectively. 
This flattening can be also seen in Table 2, and is obviously
expected if time dilation exists in light curves of 
the observed GRBs. Therefore, interpreting the power law
index and its power distribution may not be straightforward,
unless we understand how the light curve 
is stretched or even contracted.

\acknowledgments

We thank the anonymous referee for suggestions to clarify 
the earlier version of the manuscript.
We also thank William Paciesas for informative correspondence,
Hee-Won Lee and Chang-Hwan Lee for useful discussions.
We are grateful to Ethan Vishniac for hospitality while
visiting Johns Hopkins University where this work began.
This research has made use of data obtained through 
the High Energy Astrophysics Science Archive 
Research Center Online Service, 
provided by the NASA/Goddard Space Flight Center.

\clearpage

\figcaption[fig1.ps]{Averaged PDSs of the two 
subgroups of the observed GRBs with known $z$
are shown as a function of frequency 
in log-log plots. Power is in arbitrary unit 
and frequency is in Hz. Open triangles
and open squares represent far and near GRB 
subgroups, respectively. Solid
and dashed lines are the best fits of data.
 Four plots result from four different
energy bands of BATSE experiments as indicated. 
Flat components at higher frequencies $f \ga 1$ Hz show
Poisson noise.\label{fig1}}

\figcaption[fig2.ps]{Similar plots as Figure 1, 
but light curves are rescaled
by a factor of $1+z$ to remove a time dilation 
effect before calculated PDSs. \label{fig2}}

\clearpage

\begin{table}
\begin{center}
\caption{A list of the GRBs used in the analysis with 
the redshifts
and peak fluxes. The redshifts are quoted from a 
complied table in 
$http://www.aip.de/~jcg/grbgen.html$. \label{tbl-1}}
\vspace{2mm}
\begin{tabular}{cccc}
\tableline\tableline
GRB name&trigger number&redshift&peak flux \\ \tableline \tableline
GRB 000418 &  8079 & 1.118  & 1.6542 \\
GRB 991216 &  7906 & 1.02   & 91.481 \\
GRB 990510 &  7560 & 1.619  & 11.283 \\
GRB 990506 &  7549 & 1.3    & 25.122 \\
GRB 990123 &  7343 & 1.60   & 16.962 \\
GRB 980703 &  6891 & 0.966  & 2.9310 \\
GRB 980425 &  6707 & 0.0085 & 1.2451 \\
GRB 980329 &  6665 & 3.9    & 13.848 \\
GRB 971214 &  6533 & 3.42   & 2.6490 \\
GRB 970508 &  6225 & 0.835  & 1.2816 \\ \tableline \tableline
\end{tabular}
\end{center}
\end{table}

\begin{table}
\begin{center}
\caption{Obtained slopes with the least square 
fits for the two subgroups
of far and near GRBs. Fittings are repeated before and after
correction of a time dilation effect by a factor of $1+z$. \label{tbl-2}}
\vspace{2mm}
\begin{tabular}{ccccc}
\tableline\tableline
&as-it-is&&corrected& \\ \tableline
channel&far&near&far&near \\ \tableline
1&$-1.7064\pm0.115$&$-1.5436\pm0.109$&$-1.5646\pm0.149$&$-1.5082\pm0.097$ \\
2&$-1.8052\pm0.101$&$-1.5220\pm0.109$&$-1.5857\pm0.144$&$-1.4652\pm0.095$ \\
3&$-1.8811\pm0.103$&$-1.5105\pm0.095$&$-1.6008\pm0.159$&$-1.4160\pm0.090$ \\
4&$-1.3714\pm0.127$&$-1.0989\pm0.070$&$-1.3985\pm0.150$&$-1.1130\pm0.059$ \\ \tableline \tableline
\end{tabular}
\end{center}
\end{table}

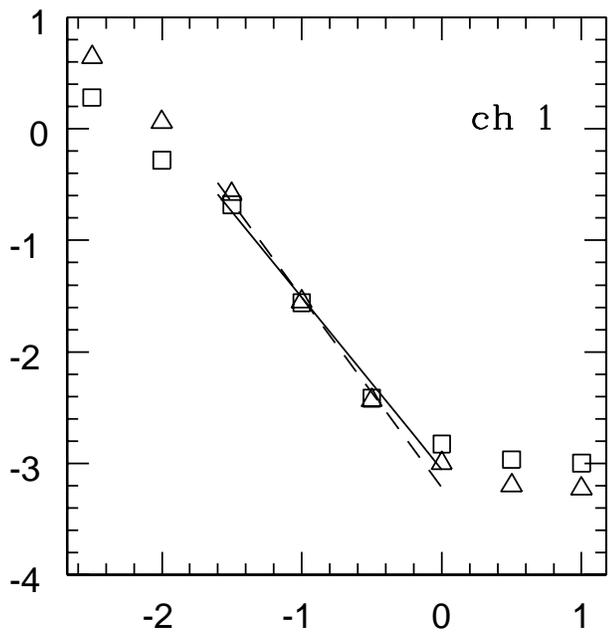
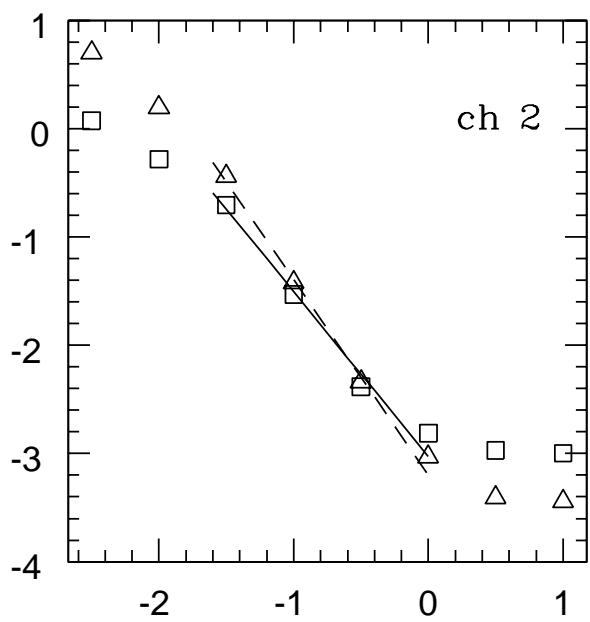
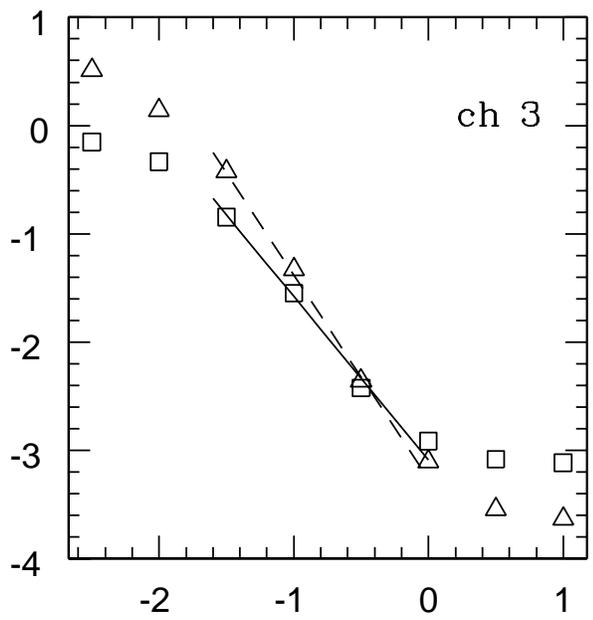
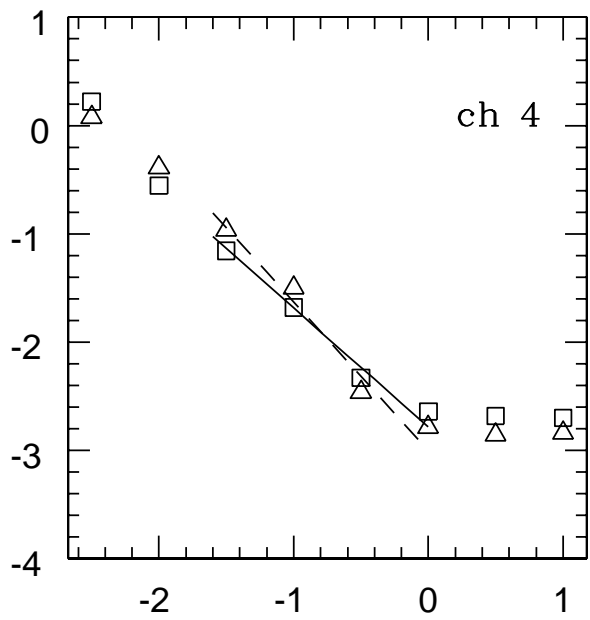
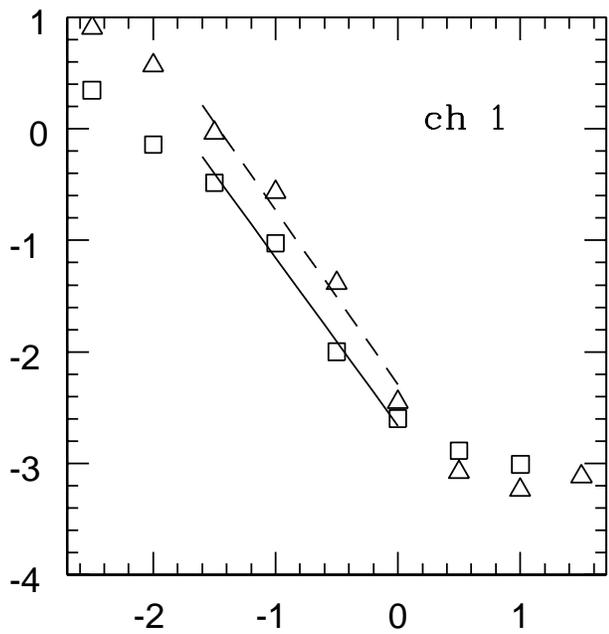
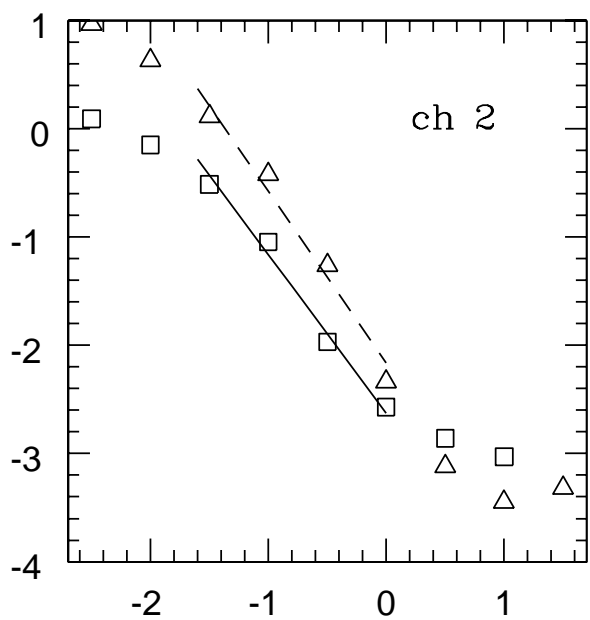
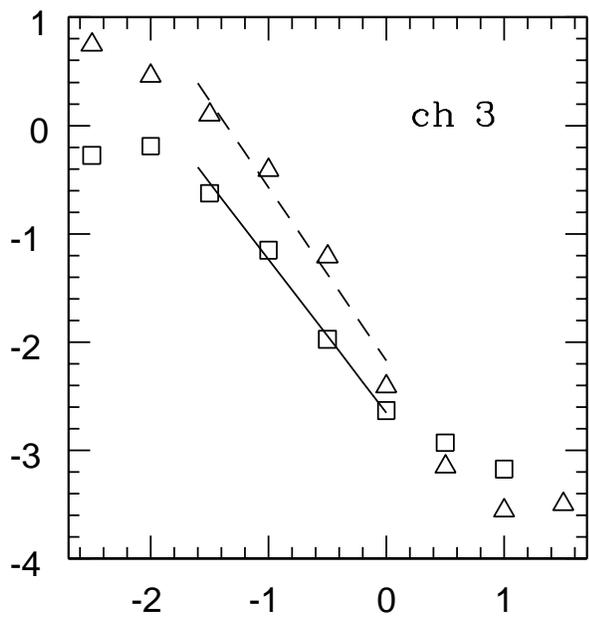
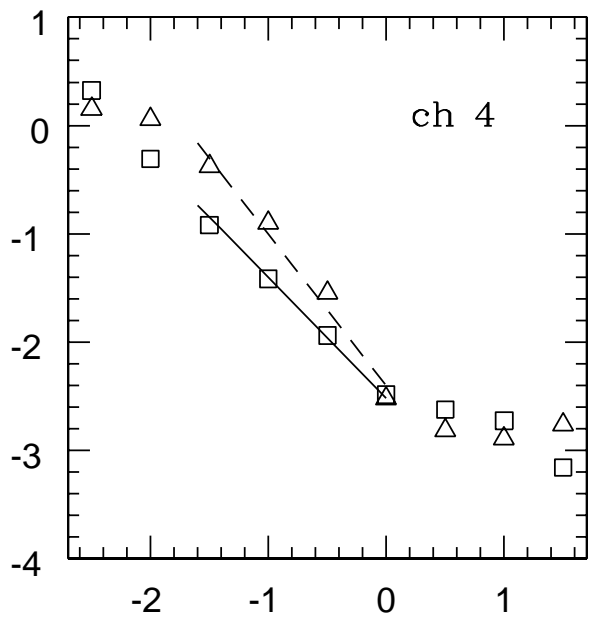
\end{document}